\documentstyle [12pt]{article}
\newcommand{\bce}{\begin{center}}
\newcommand{\ece}{\end{center}}
\newcommand{\beq}{\begin{equation}}
\newcommand{\eeq}{\end{equation}}
\newcommand{\bef}{\begin{figure}}
\newcommand{\eef}{\end{figure}}
\newcommand{\bea}{\vspace{0.25cm}\begin{eqnarray}}
\newcommand{\eea}{\end{eqnarray}}

\newcommand{\lapp}{\stackrel{<}{\sim}}
\newcommand{\gapp}{\stackrel{>}{\sim}}

\newcommand{\ba}{\begin{array}}
\newcommand{\ea}{\end{array}}

\newcommand{\etc}{{\sl etc.~}}

\newcommand{\singlespace}{
    \renewcommand{\baselinestretch}{1}\large\normalsize}
\newcommand{\doublespace}{
    \renewcommand{\baselinestretch}{1.6}\large\normalsize}

\newcommand{\footju}{\footnote[1]{ also: IKP (Theorie),
Forschungszentrum J\"ulich, D-52425 J\"ulich, F.R.G.}}
\newcommand{\footbo}{\footnote[2]{ also: ITKP, Universit\"at Bonn,
D-53115 Bonn, F.R.G.}}
\begin{document}
\pagestyle{empty}
\singlespace
\vspace{1.0in}
\begin{flushright}
November, 1993 \\
\end{flushright}
\vspace{1.0in}
\bce
{\large{\bf PION PROPERTIES IN A HOT $\pi$$N$$\Delta$ GAS}}
\vskip 1.0cm
R. RAPP\footju and J. WAMBACH{\footnotemark[1] \footbo}

\vspace{.35in}
{\it Department of Physics\\
University of Illinois at Urbana-Champaign\\
1110 West Green St., Urbana, IL 61801-3080, USA}
\ece
\vspace{.65in}
\begin{abstract}
\noindent Based on a recent meson-exchange model for the vacuum
$\pi\pi$ interaction we compute selfconsistently
the in-medium $\pi\pi$ scattering amplitude
and pion selfenergy in a hot $\pi N\Delta$ gas. The
contributions to the pion selfenergy are calculated from the $\pi\pi$
T-matrix as well as from p-wave interaction with
nucleons and thermally abundant $\Delta$'s.
Results are presented for two
scenarios believed to be realized in the relativistic
heavy ion collisions performed at the GSI-SIS and the CERN-SpS. Possible
implications for the observed soft pion enhancement at both SIS and
SpS are indicated.
\end{abstract}
\vspace{.75in}
\begin{flushright}
\singlespace
PACS Indices: 25.75.+r\\
21.65.+f\\
\end{flushright}
\newpage
\pagestyle{plain}
\baselineskip 16pt
\vskip 48pt

\newpage
\doublespace

\section{Introduction}
Although current ultrarelativistic heavy ion collisions (URHIC's),
performed e.g. at the BNL-AGS or at the CERN-SpS, presumably
have not yet produced a quark-gluon plasma (QGP),
they allow the study of hot hadronic matter
in a temperature range of typically T=100-200 MeV~\cite{StYo}.
{}From a theoretical point of view, a
precise understanding of such highly excited matter
is clearly needed to infer consequences on the equation of
state beyond the nuclear matter saturation point $\rho_0=0.16$~fm$^{-3}$
and at non-zero temperatures. From an experimental point of view,
on the other hand,
a quantitative description of the hadronic gas phase is essential
to identify signatures of a QGP, possibly formed in the early stage
of head-on collisions at the future
RHIC or LHC colliders (or maybe already with the lead beam to be
injected into the SpS soon).

\noindent
In central collisions most of the produced particles are pions.
Thus, various features of a hot interacting pion gas have been discussed
extensively in the literature~\cite{GoLe,Gavi,Shur,Sche,
ACSW,BBDS,BDSW,ChDa}.
However, even at the highest currently available energies of about
200~GeV/A in the laboratory frame, the midrapidity regions still show
an appreciable 'baryon contamination'. According to recent studies,
the $\pi\pi$ interaction in cold nuclear matter undergoes considerable
modifications already well below saturation density~\cite{CASN,MuWS};
therefore one might expect a significant impact of small baryonic
components on the pions even in the presence of a much higher
pion density. In the present paper we shall extent our selfconsistent
analysis of a thermal pion gas~\cite{RaWa} by including nucleons
as well as thermally excited $\Delta$'s. The
latter seem to be of special relevance in connection with the
'soft pion puzzle'. As was shown by Brown et al.~\cite{BrSW}, pions
arising from $\Delta$ decay give an important contribution to the
soft component of the pion-$p_T$-spectra at the AGS. Furthermore,
a recent analysis of SIS data~\cite{Meta} reports strong evidence
for so-called 'resonance matter' in which about one third of the
nucleons are excited into higher resonances (predominantly $\Delta$'s).
In such a scenario the baryons are strongly out of chemical
equilibrium.

\noindent
Our calculations are based on the assumption of
thermal equilibrium which is motivated by standard estimates for the
mean number of collisions per particle until freezeout. On the
other hand, also the pions may decouple from chemical equilibrium
prior to freezeout. This would lead to a non-zero chemical
potential $\mu_\pi$ which could be close to the pion mass~\cite{KaRu}.
Therefore we will also show some results for $\mu_\pi > 0$.

\noindent
In sect.~2 we briefly review the formalism for the
selfconsistent description of the pion gas \cite{RaWa}
which forms the framework of the present study.
In sect.~3 the model for the baryonic
contribution to the pion selfenergy
is discussed while sect.~4 gives results for the selfconsistent
pion mean field as well as the $\pi\pi$ T-matrix in the
presence of nucleons and $\Delta$'s.
We will focus on two different scenarios:
(1) 'resonance matter' characterized by temperatures below 100 MeV and
high baryon density $\rho_B\gapp 2\rho_0$ with an appreciable
$\Delta$ component (SIS conditions) and (2) SpS freezeout conditions
where $T\approx 150$ MeV and $\rho_B\lapp \rho_0$.
In sect.~5 we summarize our findings and indicate some consequences.

\section{Selfconsistent Pion Gas}
The basic ingredient for our description of the hot pion gas is the
meson-exchange model for the $\pi\pi$ interaction in free space
developed by Lohse et al.~\cite{LDHS}. It is described by a
T-matrix equation of Lippmann-Schwinger type, where -- in contrast
to ref.~\cite{LDHS} -- the Blancenbecler Sugar reduction
scheme~\cite{BbSu,PeHS} of the 4-dimensional Bethe Salpeter equation
has been employed. After partial-wave decomposition the
reduced scattering equation for
given angular momentum $J$ and isospin $I$ reads:
\bea
T_{\pi\pi}^{JI}(Z,q_1,q_2) & = & V_{\pi\pi}^{JI}(Z,q_1,q_2)
     \nonumber\\
 & &
+ \int_0^{\infty}dk \ k^2 \ 4{\omega}_k^2 \ V_{\pi\pi}^{JI}(Z,q_1,k)
 \   G_{\pi\pi}(Z,k) \ T_{\pi\pi}^{JI}(Z,k,q_2)  \ ,
\label{eq:Tmat}
\eea
where $k=|\vec k|$ \etc ; $Z$ is the total energy of the pion pair
and $G_{\pi\pi}(Z,k)$ the two-pion propagator in the CMS
frame with pions of momenta $\vec k$ and $-\vec k$ (the Lohse et al.
model also contains coupling to the $K\bar K$-channel, which has been
omitted for brevity in eq.~(\ref{eq:Tmat}) but which is included in
the calculation). In the BbS form the vacuum
two-pion propagator is given by
\beq
G_{\pi\pi}^0(Z,k)=\frac{1}{\omega_k} \  \frac{1}{Z^2-4\omega_k^2+i\eta}
\eeq
with ${\omega_k}^2=k^2+m_\pi^2$.
The pseudopotentials $V_{\pi\pi}^{JI}$ are constructed from
an effective meson Lagrangian, where t- and
s-channel $\rho$-exchange are
the dominant contributions to the s- and
p-wave interaction, respectively.
This model gives a good description of the
phase shifts and inelasticities
up to $\sim 1.5$ GeV which is more than sufficient for our purposes.

\noindent
The most important medium effect can be attributed
to a modification of the
pion propagation in the gas. At finite temperature, T, and pion
chemical potential, $\mu_\pi$, the in-medium single-pion propagator is
given by
\beq
D_\pi(\omega,k;\mu_\pi,T )=\lbrack \omega^2-m_\pi^2-k^2-
\Sigma_\pi(\omega,k;\mu_\pi,T) \rbrack ^{-1}  \ ,
\eeq
where $\Sigma_\pi$ denotes the pion selfenergy
in the pion gas. Invoking
the quasiparticle approximation (QPA),
i.e. expanding $\Sigma_\pi$ around
the 'quasiparticle pole'
\beq
e_k=[\omega_k^2+Re\Sigma_\pi(e_k,k;\mu_\pi,T)]^{1/2}
\eeq
and retaining only terms up to first order,
\beq
\Sigma_\pi(\omega,k)\approx \Sigma_\pi(e_k,k)+
\frac{\partial \Sigma_\pi(\omega,k)}{\partial\omega^2}|_{e_k}
 \ (\omega^2-e_k^2) \ ,
\eeq
enables one to calculate the in-medium two-pion propagator
\beq
G_{\pi\pi}(Z,k;\mu_\pi,T)=\int \frac{i dk_0}{2\pi} \
D_\pi(k_0,\vec k;\mu_\pi,T) \ D_\pi(Z-k_0,-\vec k;\mu_\pi,T)
\eeq
analytically. In addition, taking into account the Bose statistics
of the surrounding gas, leads to the result
\beq
G_{\pi\pi}(Z, k;\mu_\pi,T)=
\frac{1}{\bar \omega}_k \ \frac{z_k^2 \ (1+2f^\pi(e_k;\mu_\pi,T))}
{Z^2-4{\bar \omega}_k^2}
\eeq
with
\bea
z_k & \equiv & (1-\frac{\partial \Sigma_\pi(\omega,k)}
{\partial\omega^2}
|_{e_k})^{-1} \quad {\rm the \ pole strength} \ ,
\nonumber\\
\bar \omega_k^2 & \equiv & e_k^2 + i \ z_k \
Im\Sigma_\pi(e_k,k) \quad
{\rm quasi \ pion \ dispersion \ relation} \  ,
\nonumber\\
f^\pi(e_k;\mu_\pi,T) & = & (\exp \lbrack
(e_k-\mu_{\pi})/T \rbrack -1)^{-1}
 \ {\rm thermal \  Bose \  factor} \ .
\eea
The pion selfenergy, on the other hand, is obtained from the on-shell
forward scattering  T-matrix as~\cite{RaWa}
\beq
\Sigma_\pi(\omega,k;\mu_\pi,T)=\frac{\pi}{k} \int\limits_0^\infty dp \
\frac{p}{e_p} \ f^\pi(e_p;\mu_\pi,T) \int
\limits_{E_{min}}^{E_{max}} dE_{cms} \ E_{cms}^3 \
T_{\pi\pi}(E_{cms}) \ .
\label{eq:Sig2}
\eeq
Eqs. (1), (7) and (9)
define a selfconsistency problem of Brueckner type which can be solved
iteratively.

\section{Pion Selfenergy in Hot Nuclear Matter}
Nuclear matter, at finite temperature, not only consists of nucleons but
-- due to excitations via N-N collisions -- also contains an admixture
of baryonic resonances. Because of its relatively
small excitation energy
and high statistical degeneracy the most abundant resonance is the
$\Delta(1232)$ which we will include in our gas scenario. Thus the total
baryon density at given temperature $T$ can be expressed as
\bea
\rho_{B}(\mu_N,\mu_\Delta,T) & = & \rho_N(\mu_N,T)+
\rho_\Delta(\mu_\Delta,T)
\nonumber\\
& = & \int\frac{d^3q}{(2\pi)^3} \
[4f^N(q;\mu_N,T)+16f^\Delta(q;\mu_\Delta,T)]
\eea
with $f^a(q;\mu_a,T)=(\exp[(E_q^a-\mu_a)/T]+1)^{-1}$ and
$E_q^a=\sqrt{q^2+M_a^2} \ $  for $a=N,\Delta$.
Assuming chemical equilibrium
($\mu_N=\mu_\Delta$), at $T=100$ MeV,
every fourth nucleon is excited into
a $\Delta$, whereas for $T\ge170$ MeV $\Delta$'s are
already in the majority (see Fig.~1).

\noindent To calculate the pion selfenergy $\Sigma_{N\Delta}$ in the
$N\Delta$ gas,  we use the finite-temperature extension of
the standard particle-hole model~\cite{ErWe}. In this
approach, the most important modification of the pion propagation is
attributed to p-wave excitations
of the type $\alpha=ab^{-1}$ with $a,b=N,\Delta$.
The amplitude for such processes is obtained
from a folding integral over
the particle and hole propagator resulting in the standard Lindhard
functions:
\beq
\phi_\alpha(\omega,k)=i \int \frac{d^4p}{(2\pi)^4} \
[G_a^0(p_0+\omega,\vec p +\vec k)
\  G_{b^{-1}}^0(p_0,\vec p) + G_a^0(p_0-\omega,\vec p -\vec k) \
G_{b^{-1}}^0(p_0,\vec p) ]
\eeq
with
\bea
G_a^0(p_0,\vec p) & = & \frac{1-f^a(p)}{p_0-E_a(p)+i\eta} \
\mbox{the nucleon or}
 \  \Delta  \ \mbox{propagator}  ,
\nonumber\\
G_{b^{-1}}^0(p_0,\vec p) & = & \frac{f^b(p)}{p_0-E_b(p)-i\eta} \
\mbox{the nucleon-hole or} \ \Delta \mbox{-hole propagator} \ ,
\eea
where $E_a(p)=E^a_p-M_b-\frac{i}{2} \Gamma_a \ , \
E_b(p)=E_p^b-M_b+\frac{i}{2} \Gamma_b$ .
For the $\Delta$ width, $\Gamma_\Delta$, we choose the relativistic
parameterization of ref.~\cite{ErWe}
supplemented by a density dependent
term to account for higher-order medium corrections. Then
\beq
\Gamma_\Delta(k;\rho_B)=\frac{2}{3} \ \frac{f_{\pi N\Delta}^2}{4\pi}  \
\frac{q_{cms}^3}{\sqrt{s}} \ \frac{M_N}{m_\pi^2} \
\Gamma_\pi^2(k) \ F_{fit}
+ 20 \ \frac{\rho_{B}}{\rho_0} {\rm MeV}
\eeq
with $q_{cms}$ being the pion momentum in
the $\pi N$ CMS and $\sqrt{s}$ the $\pi N$
CMS energy. The form factor $\Gamma_\pi(k)=
(\Lambda_\pi^2-m_\pi^2)/(\Lambda_\pi^2+k^2)$
accounts for the hadronic size of the pion-baryon vertex
($\Lambda_\pi=1200$ MeV).
The density-dependent term has been estimated from
pion-nucleus optical potentials. In
the same spirit we take into account a width for the nucleon
$\Gamma_N=10\rho_B/\rho_0$ MeV. An additional factor
$F_{fit}=1.36$ has been introduced in order to reproduce the free space
value $\Gamma_\Delta= 115$ MeV at the
resonant momentum $k_{res}=297$ MeV/c.

\noindent The second term in eq.~(11) is obtained from the first one by
replacing $(\omega,\vec k) \rightarrow (-\omega,-\vec k)$.
It characterizes
the exchange graph which is necessary to ensure Bose symmetry when
interchanging the in- and outgoing pion line. After a standard contour
integration  in the complex $p_0$ plane the Lindhard functions can be
simplified as
\beq
\phi_\alpha(\omega,k)=-\int dp \frac{p^2 \
f^b(p)}{(2\pi)^2} \int\limits_{-1}^{+1}
 \ dx \ \sum_{m=1}^{2} \ \frac{1-f_{pk}^a(x)}
 {\pm\omega+E_p^b-E_{pk}^a(x)+
 \frac{i}{2}(\Gamma_a+\Gamma_b)}   \
\eeq
with  $E_{pk}^a(x)=(M_a^2+p^2+k^2+2pkx)^{1/2}$, $f_{pk}^a(x)=
(1+\exp[(E_{pk}^a(x)-\mu_a)/T])^{-1}$ . The $\pm$-sign corresponds to
$m=1,2$, respectively. From the Lindhard functions we calculate the
so-called pionic susceptibilities in lowest order as
\beq
\chi_\alpha^{(0)}(\omega,k)={\left ( \frac{f_{\pi\alpha} \
\Gamma_\pi(k)} {m_\pi}
\right )}^2  \ SI(\alpha) \ \phi_\alpha(\omega,k) \ .
\eeq
The values for the spin-isospin transition
factors $SI(\alpha)$ and for the
pion-baryon coupling constants $f_{\pi\alpha}$
are summarized in table 1.
In the latter, the following relations have been used: $f_{\pi N\Delta}=
2f_{\pi NN}$ (Chew-Low factor~\cite{ChLo}) and $f_{\pi\Delta\Delta}=
\frac{1}{5} f_{\pi NN}$ (from the constituent quark model~\cite{BrWe}).
In evaluating the pion selfenergy it is crucial to include
the short-range spin-isospin correlations between particle and hole.
These correlations can also induce transitions between the various
excitation modes. Therefore, using Migdal's approximation, one
ends up with a system of 4 coupled equations
in the fully dressed susceptibilities
$\chi_\alpha$:
\beq
\chi_\alpha=\chi_\alpha^{(0)}-\sum_{\beta} \ \chi_\alpha^{(0)}  \
{g'}_{\alpha\beta} \ \chi_\beta
\eeq
with $\alpha, \beta= NN^{-1}, \Delta N^{-1},
N\Delta^{-1}, \Delta\Delta^{-1}$ .
For the Migdal parameters ${g'}_{\alpha\beta}$
we choose momentum independent
values ${g'}_{NN}=0.75$, ${g'}_{N \Delta}={g'}_{\Delta\Delta}=0.33$.
The value for ${g'}_{NN}$ can be deduced from Gamow-Teller resonance
systematics~\cite{Oste}; all others are
less explored, and we fix them at the classical
Lorentz-Lorenz value of 0.33.
Eq.~(16) is then solved by matrix inversion,
\beq
\vec \chi = A_{int}^{-1} \ \vec\chi^{(0)} \ ,
\eeq
where we have defined the $4\times 4$ interaction matrix
\beq
(A_{int})_{\alpha\beta}\equiv \delta_{\alpha\beta} + \chi_\alpha^{(0)} \
{g'}_{\alpha\beta}  \ .
\eeq
The total susceptibility is the sum over the 4 channels:
\beq
\chi_{N\Delta}=\sum_{\alpha} \ \chi_\alpha \ ,
\eeq
and the pion selfenergy is finally given by
\beq
\Sigma_{N\Delta}(\omega,k;\mu_N,\mu_\Delta,T)= -k^2 \ z_\pi^2 \
\chi_{N\Delta}(\omega,k;\mu_N,\mu_\Delta,T)  \ .
\eeq
Here a renormalization constant $z_\pi$ has been introduced to
account for all processes other than the explicitly treated p-wave
excitations. Migdal et al.~\cite{MSTV} have estimated it in a dilute gas
approximation to be
\beq
z_\pi=(1+4\pi\lambda\rho_B)^{-1/2} \ ,
\eeq
where $\lambda=0.036$~fm$^{-3}$.

\noindent The on-shell pion selfenergy in the $N\Delta$ gas is now
obtained from a selfconsistent solution of the in-medium pion dispersion
relation
\beq
e_k^2=\omega_k^2+\Sigma_{N\Delta}(e_k,k)
\eeq
with $e_k^2=\omega_k^2+Re\Sigma_{N\Delta}(e_k,k)$,
i.e. the selfconsistency
is restricted to the real part of the quasi pion dispersion relation.

\noindent The results for the pion mean field
(optical potential), defined as
\beq
U_{N\Delta}(k)\equiv \frac{\Sigma_{N\Delta}(e_k,k)}{2\omega_k} \ ,
\eeq
are displayed in Fig.~2 at various temperatures
and total baryon densities.
The relative abundances of nucleons and $\Delta$'s were fixed via the
condition for chemical equilibrium, i.e. $\mu_N=\mu_\Delta$.
The real part of $U_{N\Delta}$ (upper panels in Fig.~2) shows a
characteristic minimum induced by resonant $\Delta N^{-1}$
excitations. The latter also show up in the imaginary part exhibiting
a rather pronounced peak at the resonant
pion momentum of about 300 MeV/c
(lower panels).
At higher temperatures the increased thermal
motion of the nucleons leads
to a slight downward shift of the peak and we
also observe an overall reduction in
the absorption, which implies a decrease
in $Re U_{N\Delta}$ as well. This
effect is directly related to the relative
decrease of the nucleon component
in the gas (see also Fig.~1). Thus the
resonant $\Delta N^{-1}$ excitations
are suppressed, partly because there are relatively fewer nucleons and
partly due to stronger Pauli blocking
of possibly excited $\Delta$'s. On
the other hand, the pion interaction with
the thermally abundant $\Delta$'s
is essentially non-resonant.

\noindent We would also like to mention that the selfconsistent
solution of eq.~(22) is not always unique.
Starting from nucleon densities
$\rho_N\approx 1.5\rho_0$ (slightly depending on
temperature and $\Delta$
density) the $\Delta N^{-1}$ branch, and eventually also the $NN^{-1}$
branch, begin to mix so strongly that
additional solutions appear. Physically this can be understood as an
explicit splitting of the pion propagation into different modes.

\section{The $\pi N\Delta$  Gas}
We now extend the formalism described in sect.~2 for a pure pion gas
by including the baryonic selfenergy
contribution $\Sigma_{N\Delta}$. Then
the total pion selfenergy receives two contributions:
\beq
\Sigma_{\pi N\Delta}(\omega,k;\mu_\pi,\mu_N,\mu_\Delta,T) =
\Sigma_\pi(\omega,k;\mu_\pi,T)+\Sigma_{N\Delta}
(\omega,k;\mu_N,\mu_\Delta,T) \ .
\eeq
Consequently, in eqs.~(3)-(8), $\Sigma_\pi$ has to be replaced by
$\Sigma_{\pi N\Delta}$, such that
\beq
G_{\pi\pi}(Z,k;\mu_\pi,\mu_N,\mu_\Delta,T)=\frac{1}{\bar{\omega}_k} \
\frac{z_k^2 \ (1+2f^\pi(e_k;\mu_\pi,T))}{Z^2-4\bar{\omega}^2_k}
\eeq
with
\bea
z_k & = & (1-\frac{\partial\Sigma_{\pi N\Delta}}
{\partial\omega^2}|_{e_k})^{-1} \ ,
\nonumber\\
\bar{\omega}^2_k & = & e_k^2+iz_k \ Im\Sigma_{\pi N\Delta}(e_k,k) \ ,
\nonumber\\
e_k^2 & = & \omega_k^2+Re\Sigma_{\pi N\Delta}(e_k,k)  \  .
\eea
Therefore eqs.~(1), (9), (20) and (25) form our extended set of
selfconsistent equations, which we solve by numerical iteration.
The pion density $n_\pi(\omega_k;\mu_\pi^{(0)},T)$ is kept fixed
by readjusting $\mu_\pi$ after each iteration step according to the
change in the quasi pion energy $e_k$ such that, after convergence,
$n_\pi(\omega_k;\mu_\pi^{(0)},T)=n_\pi(e_k;\mu_\pi^*,T)$ with the
appropriate value for $\mu_\pi^*$.

\noindent Let us first concentrate on a scenario which may be realized
at SIS energies. A recent compilation of data~\cite{Meta} gives
strong evidence for 'resonance matter', i.e. hot nuclear matter
with an abundance of nucleonic resonances
in excess of values obtained for
chemical equilibrium; at SIS temperatures of typically $T=75$ MeV the
relative abundance of $\Delta$'s in
chemically equilibrated nuclear matter
($\mu_N=\mu_\Delta$) would be about 10\% (see also Fig.~1); however, the
analysis of various data in conjunction with BUU calculations suggests
a nucleon-to-delta ratio of about 2:1~\cite{Meta}. Fig.~3
shows our selfconsistent results for the on-shell $\pi\pi$ T-matrix
$T_{\pi\pi}(Z)$ (upper panels) as well as the
pion optical potential $U_{\pi N\Delta}(k)$ (middle panels)
at $T=75$ MeV, $\rho_B=2\rho_0$
and for various nucleon-to-delta ratios (fixed by independent
choice of $\mu_N$ and $\mu_\Delta$).
For the pion chemical potential we
choose $\mu_\pi^{(0)}=0$. At $T=75$ MeV
this corresponds to a pion density of 0.01~fm$^{-3}$
which is too small to
show any visible effect on $T_{\pi\pi}$ or
$U_{\pi N\Delta}$. The optical
potential is  dominated by the $\Delta N^{-1}$
excitation channel. Due to
the selfconsistency requirement the lowering of the quasi pion energy
leads to an upward shift of the maximum in $Im U_{\pi N\Delta}$
which now peaks above the resonant momentum of
$k=297$ MeV/c in free space.
As in the previous section,
we see that an increase of the $\Delta$ abundance at fixed total
baryon density reduces the magnitude of the pion mean field.
The in-medium $\pi\pi$ T-matrix in the
$\sigma$-channel (upper-left part
in Fig.~3) as well as the $\rho$-channel
(upper-right part) shows a strong
reduction of the peak values as compared to the vacuum. The depletion of
strength in the $\sigma$-channel in the
energy range 400 MeV $\leq Z\leq$ 800 MeV
comes from the rapid variation in $Re U_{\pi N\Delta}$ for pion momenta
$k\approx 250-400$ MeV/c, turning from strong attraction to repulsion,
thus suppressing pion modes of energies
$e_k\approx 200-400$ MeV. In the
same momentum region, the real part of
${\partial\Sigma_{\pi N\Delta}}/{\partial \omega^2}|_{e_k}$ acquires
large positive values (lower-left panel in Fig.~3) which numerically
further enhances the depletion effect.
This mechanism is also responsible
for strength accumulation slightly below the $K\bar K$ threshold at
$Z\approx 970$ MeV.
We furthermore observe considerable accumulation of strength below the
two-pion threshold. To decide on whether
two-pion bound states are formed
requires an analysis beyond the quasiparticle approximation (QPA)
taking into account the off-shell behavior of
the pion selfenergy in more detail~\cite{aoui}. Besides broadening, the
$\rho$-resonance (upper-right panel) shows a
considerable upward shift of
about $70-90$ MeV. The same qualitative features
have also been found for
cold nuclear matter in  ref.~\cite{MuWS} where a similar ansatz for the
pion selfenergy has been used neglecting Fermi motion, however.
Furthermore, the 2-pion propagator was calculated
in a three-branch model,
whereas in the present  paper the quasiparticle
picture is employed. The
validity of the QPA is limited to rather small pion widths,
$\Gamma_k=-2 Im U_{\pi N\Delta}(k)$. It may therefore
be questioned for nucleon densities above
$\sim $1.5$\rho_0$ since the ratio
$\Gamma_k/e_k$ is already close to one in the peak region,
$k\approx 300 $ MeV/c. However, we have checked the sensitivity of our
results with respect to different orders in the expansion of the pion
selfenergy by performing calculations with purely real
pole strength, i.e neglecting the imaginary part of
$\partial\Sigma_{\pi N\Delta}/\partial\omega^2|_{e_k}$. It turns
out that these results coincide with those for complex
$z_k$ within a few percent -- in contrast to the
pure pion gas analysis of ref.~\cite{RaWa} where a strong dependence on
$Im{\partial\Sigma_\pi}/\partial\omega^2|_{e_k}$ was found. Thus one
might be encouraged to conclude that the QPA  produces meaningful
results even at baryon densities as high as $\rho_B\approx2\rho_0$.

\noindent The second scenario which we would like to discuss is believed
to be realized in the midrapidity regions of the SpS
experiments at CERN.
In central collisions at an energy of about 200~GeV/A in the lab frame,
the two colliding nuclei almost pass through each other while depositing
large amounts of energy in the central zone.
After hadronization one should
then encounter a hot and dense gas of mesons contaminated with a net
baryon density of about $0.5\rho_0$ as extracted from measured
multiplicities~\cite{Naga}. The predominant meson species is
the pion. At a temperature of $T=150$ MeV and in chemical equilibrium
($\mu_\pi=\mu_K=\mu_{\bar{K}}=0$) the kaon-/ antikaon
density is already less
than $\frac{1}{6}$ of the pion density. For a hot $\pi N\Delta$ gas
at $T=150$ MeV we again perform selfconsistent calculations
as described in the beginning of this section. This time, the gas is
assumed to be in chemical equilibrium characterized by the condition
\beq
\mu_\pi^{(0)}+\mu_N=\mu_\Delta  \  .
\eeq
Thus we end up with three independent thermodynamic parameters: the
temperature $T$, the pion density $n_\pi(\mu_\pi^{(0)},T)$ -- which,
at a given temperature, is fixed by a starting value of
the pion chemical potential -- and
the total baryon density $\rho_B$. At given $T$ and $\mu_\pi^{(0)}$ the
relative abundances of nucleons and $\Delta$'s
are then determined  from eq.~(27).
The results for the in-medium $\pi\pi$ T-matrix and the pion optical
potentials are displayed in Fig.~4 for
different values of $\mu_\pi^{(0)}$
and $\rho_B$. Even at rather low baryon densities of $\rho_B=0.5\rho_0$
and at $\mu_\pi^{(0)}=0$ one clearly recognizes the signatures of
the baryonic selfenergy contribution (which are even more pronounced
for $\rho_B=1.0\rho_0$): the $\Delta$-peak in $Im U_{\pi N\Delta}$ at
$k\approx 300$ MeV/c (lower-middle part in Fig.~4)
and a corresponding minimum in $Re U_{\pi N\Delta}$
at $k\approx 200$ MeV/c (lower-left part)
as well as an appreciable reduction of the T-matrix peak values in both
the $\sigma$- and $\rho$-channel (upper-left and -middle part).
However, the attraction in $Re U_{\pi N\Delta}$ is much
less than under SIS conditions. As a consequence two-pion quasi bound
states in the s-wave now appear close to the two-pion threshold.
The impact of the pionic gas
component becomes apparent, if we increase $\mu_\pi^{(0)}$ from zero to
$100$ MeV which is equivalent to an increase in pion density from
0.12~fm$^{-3}$ to 0.27~fm$^{-3}$. The optical potentials now exhibit
considerable non-zero values at zero momentum.
In $Im U_{\pi N\Delta}$ this is essentially caused
by the s-wave $\pi\pi$ interaction which is enhanced
due to large accumulation of strength in
the threshold region of $Im T^{00}_{\pi\pi}$
(upper-left part of Fig.~4).
This threshold enhancement is a consequence of the
Bosefactors $(1+2f^\pi)$
in $G_{\pi\pi}$ leading to a stronger weighting of small momenta. The
downward shift of the in-medium two-pion
threshold is due to the decrease
in the pion mass of nearly 20 MeV,
generated by $Re U_{\pi N\Delta}(k=0)$.
At $\mu_\pi^{(0)}=100$ MeV,
the resonances in the different $JI$-channels
show the remarkable feature
that they seem to accumulate the major part of the strength in the
corresponding partial wave. In the s-wave this is the
case for the $2\pi$ bound
state, the resonant structure
near the $K\bar K$ threshold as well as the
$\epsilon (1400)$ (which is now shifted to $Z\approx 1150$ MeV). In the
p-wave the $\rho$ peak becomes strongly enhanced
(upper-middle part of Fig.~4)  and the same is observed for
the d-wave $f_2(1270)$ in $Im T^{20}_{\pi\pi}$
(upper-right part of Fig.~4).
In  ref.~\cite{RaWa} it was shown that these enhancements are related to
the energy dependence of the absorption (lower-right part of Fig.~4).

\noindent
{}From the partial-wave decomposed T-matrices we can compute the
in-medium total cross sections for $\pi^+\pi^-$ scattering as well as
for the charge-exchange reaction $\pi^+\pi^-\rightarrow \pi^0\pi^0$.
This is done by a straightforward
transformation of the T-matrix from the
isospin to the particle basis~\cite{LDHS} :
\bea
\langle\pi^+\pi^-|T^J_{\pi\pi}|\pi^+\pi^-\rangle & = & \frac{2}{3}
\ T^{J0}_{\pi\pi}  + T^{J1}_{\pi\pi} + \frac{1}{3} \ T^{J2}_{\pi\pi} \ ,
\nonumber\\
\langle\pi^0\pi^0|T^J_{\pi\pi}|\pi^+\pi^-\rangle & = &
\frac{\sqrt 2}{3} \ T^{J0}_{\pi\pi}-
\frac{\sqrt 2}{3} \ T^{J2}_{\pi\pi}   \ .
\eea
The total cross section is then obtained as
\beq
\sigma_{tot}^{\pi^+\pi^-\rightarrow \pi \pi}(Z)=\frac{\pi^3}{4} \ Z^2 \
\sum_J \ (2J+1) \
|\langle\pi \pi|T^J_{\pi\pi}(Z)|\pi^+\pi^-\rangle|^2  \  ,
\eeq
where the final state consists of either
two charged or two neutral pions.
Fig.~5 summarizes our cross section results for the CERN conditions.
For both charged and neutral pions in the final state the in-medium
cross section is enhanced over the vacuum results for $Z\ge 800$ MeV as
soon as $\mu_\pi^{(0)}$ reaches about $m_\pi/2$ (at $T=150$ MeV,
$\rho_B=0.5\rho_0$). Especially the $f_2(1270)$-resonance seems to be
quite sensitive to this enhancement mechanism, which might show up in
future dilepton or two-photon measurements in form of a resonance
narrowing. On the other hand, there is a suppression of
$\sigma^{in-medium}_{tot}$
in the intermediate energy range 300 MeV $\le Z \le$ 700 MeV
where, in our model, there is no genuine $\pi\pi$ resonance. For
chemically equilibrated pions at $T=150$ MeV
this suppression can be quite
large, especially at higher $\rho_B$. A similar
suppression is found, in
this case, for the $\rho$ peak region (upper part of Fig.~5).

\section{Summary}
Starting from a realistic model for the
$\pi\pi$ interaction in free space
we have examined medium modifications of the $\pi\pi$ T-matrix and the
single-pion  dispersion relation in a hot gas of pions, nucleons and
$\Delta$'s. In evaluating the in-medium $\pi\pi$ T-matrix we have taken
into account statistical (Bose factors)  as well as dynamical
(pion selfenergy) effects on the $2\pi$ propagator in the
scattering equation. To simplify the numerical treatment we have
applied the QPA. The pion gas contribution
to the selfenergy was calculated from
the $\pi\pi$ T-matrix while, for the
baryonic component, a standard treatment using finite-temperature
Lindhard functions
has been adopted. We have solved the resulting selfconsistency
problem by numerical iteration.
In view of possible applications we have focused on two
different scenarios corresponding to the URHIC's performed
at the GSI-SIS and at the CERN-SpS.

\noindent
For the SIS scenario, characterized by baryon densities
$\rho_B\approx 2-3 \rho_0$ with about $1/3-1/2$ of the nucleons excited
into $\Delta$'s  and by very low pion density, we find
a strong reduction in the imaginary part
of the in-medium $\pi\pi$ T-matrix.
The pion optical potential exhibits considerable attraction for pion
momenta $k\approx 200-300$ MeV/c which is generated by resonant
$\Delta N^{-1}$ excitations. Increasing the relative
$\Delta$ abundance is shown to reduce this attraction somewhat
because of a corresponding decrease of the nucleon abundance at fixed
$\rho_B$. As was recently pointed out by Koch and
Bertsch~\cite{KoBe}, such an attraction is a promising
candidate to explain (at least partially) the
low-$p_T$ enhancement in the
pion $p_T$-spectra observed at SIS: assuming an adiabatically expanding
hadron gas, the heavier nucleons move at a much
smaller velocity than the
pions;  thus even low-momentum pions may 'escape' from the nucleonic
component of  the fireball, in which case the above
mentioned attraction should lead to a significant softening.

\noindent
The SpS scenario, on the other hand, is characterized by much higher
temperatures of $T\approx 150$ MeV  and relatively
small baryon densities
$\rho_B\lapp \rho_0$. As a consequence, the attraction in the pion mean
field is much  smaller than in the high
baryon density case, even though
some of it is restored by the increase in pion density. This additional
attraction, however, is not able to account
for the observed pion low-$p_T$ enhancement in
the SpS experiments~\cite{KoBe}. As long as we assume the
pions to be in chemical equilibrium, the imaginary part of the $\pi\pi$
T-matrix is reduced overall as compared to the vacuum case, although
less pronounced than under SIS conditions. For finite pion chemical
potentials, $\mu_\pi\gapp m_\pi/2$, the various partial-wave channels
of the T-matrix become enhanced over the free space results, however,
especially in vicinity of the resonances. As suggested by numerical
simulations of the bosonic Boltzmann equation~\cite{BDSW}, such an
enhancement could probably lead to thermalization of the pionic gas
component and hence generate an excess of low-$p_T$ pions through the
$(1+f^\pi)$ factors in the collision integral. One should be cautious,
however, in trusting the QPA beyond $T=150$ MeV and for high pion
chemical potentials, since the off-shell properties of $\Sigma_\pi$
(especially in the imaginary part) become important.

\vspace{2cm}
\bce
{\bf Acknowledgement}
\ece
\vspace{0.5cm}

\noindent
We thank D. Blaschke, G. Chanfray, G. R\"opke, P. Schuck and
J. Speth for useful discussions.
This work was supported in part by a grant from the National Science
Foundation, NSF-PHY-89-21025.
\vfill\eject

\vfill\eject
\vspace{10cm}
\bce
{\bf TABLES}
\ece
\vskip 2cm
\begin{table}[h]
\begin{center}
\begin{tabular}{|c|c|c|c|c|}
\hline
$\alpha$ & $NN^{-1}$ & $\Delta N^{-1}$ &
$N\Delta^{-1}$ & $\Delta\Delta^{-1}$ \\
\hline
$SI(\alpha)$ & 4 & 16/9 & 4/9 & 400 \\
\hline
$f_{\pi\alpha}^2/4\pi$ & 0.081 & 0.324 & 0.324 & 0.00324 \\
\hline
\end{tabular}
\end{center}
\caption{\em Spin-isospin transition factors and coupling constants for
pion induced p-wave particle-hole excitations in a hot $N\Delta$ gas.}
\end{table}

\pagebreak
%
%
\bce
{\bf FIGURE CAPTIONS}
\ece
\vskip 2.0cm
\begin{itemize}
\item[{\bf Fig.~1}:] Relative abundances of nucleons (full line)
and $\Delta$'s (dashed line) in
chemical equilibrium ($\mu_N=\mu_\Delta$) at a total
baryon density of $\rho_B=\rho_0$.

\vskip 0.5cm

\item[{\bf Fig.~2}:] Selfconsistent pion mean fields in a hot
$N\Delta$ gas in chemical equilibrium ($\mu_N=\mu_\Delta$)
at two different total baryon densities:  \\
upper part: real part of the pion mean field; \\
lower part: imaginary part of the pion mean field; \\
(long-dashed lines: $T=100$ MeV, short-dashed lines: $T=150$ MeV,
dot\-ted lines: $T=200$ MeV).

\vskip 0.5cm

\item[{\bf Fig.~3}:] Selfconsistent $\pi N\Delta$ gas at SIS
conditions ($T=75$ MeV, $\rho_B=2\rho_0$, $\mu_\pi^{(0)}=0$): \\
upper part: on-shell $\pi\pi$ T-matrix in $\sigma$-channel (JI=00)
and $\rho$-channel (JI=11); \\
middle part: real and imaginary part of the pion mean field;  \\
lower part: real and imaginary part of the energy derivative
of $\Sigma_{\pi N\Delta}$;  \\
(long-dashed lines: nucleon-to-delta ratio N:$\Delta$=4:1, short-dashed
lines: N:$\Delta$=2:1, dotted lines:
N:$\Delta$=1:1; the full lines in the
upper part correspond to the $\pi\pi$ T-matrix in free space).

\vskip 0.5cm

\item[{\bf Fig.~4}:] Selfconsistent $\pi N\Delta$ gas at CERN
conditions ($T=150$ MeV, $\mu_N+\mu_\pi^{(0)}=\mu_\Delta$):  \\
upper part: on-shell $\pi\pi$ T-matrix in $\sigma$-channel (JI=00),
$\rho$-channel (JI=11) and $f_2$-channel (JI=20);  \\
lower part: pion mean field and imaginary part of the energy derivative
of $\Sigma_{\pi N\Delta}$;  \\
(long-dashed lines: $\mu_\pi^{(0)}=0$ and $\rho_B=0.5\rho_0$,
short-dashed lines: $\mu_\pi^{(0)}=0$ and $\rho_B=\rho_0$,
dotted lines: $\mu_\pi^{(0)}=100$ MeV and $\rho_B=0.5\rho_0$;
the full lines in the upper part correspond to the $\pi\pi$ T-matrix
in free space).

\vskip 0.5cm

\item[{\bf Fig.~5}:] $\pi\pi$ cross sections
in a $\pi N\Delta$ gas at CERN
conditions ($T=150$ MeV, $\mu_N+\mu_\pi^{(0)}=\mu_\Delta$):  \\
upper part: $\pi^+\pi^-\rightarrow \pi^+\pi^-$ scattering;  \\
lower part: $\pi^+\pi^-\rightarrow \pi^0\pi^0$
charge-exchange reaction; \\
(line identification as in upper part of Fig.~4
except that the dotted lines are now for $\mu_\pi^{(0)}=75$ MeV).
\end{itemize}


\begin{thebibliography}{99}
\bibitem{StYo} J. Stachel and G. R. Young, Ann. Rev. Nucl. Part. Sci.
{\bf 42} (1992) 537.
\bibitem{GoLe} J. L. Goity and H. Leutwyler, Phys. Lett. {\bf B228}
(1989) 515.
\bibitem{Gavi} S. Gavin, Nucl. Phys. {\bf B351} (1991) 561.
\bibitem{Shur} E.V. Shuryak, Nucl. Phys. {\bf A533} (1991) 761.
\bibitem{Sche} A. Schenk, Nucl. Phys. {\bf B363} (1991) 97.
\bibitem{ACSW} Z. Aouissat, G. Chanfray, P. Schuck and G. Welke,
Zeit. Phys. {\bf A340} (1991) 347.
\bibitem{BBDS} H.W. Barz, G.Bertsch, P.Danielewicz and H. Schulz,
Phys. Lett. {\bf B275} (1992) 19.
\bibitem{BDSW} H.W. Barz, P. Danielewicz, H. Schulz and G. Welke, Phys.
Lett. {\bf B287} (1992) 40.
\bibitem{ChDa} G. Chanfray and D. Davesne, to be published.
\bibitem{CASN} G. Chanfray, Z. Aouissat, P. Schuck and W. N\"orenberg,
Phys. Lett. {\bf B256} (1991) 325.
\bibitem{MuWS} V. Mull, J. Wambach and J. Speth, Phys. Lett. {\bf B286}
(1992) 13.
\bibitem{RaWa} R. Rapp and J. Wambach, Phys. Lett {\bf B315} (1993) 220.
\bibitem{BrSW} G. E. Brown, J. Stachel and G. Welke, Phys. Lett.
{\bf B253} (1991) 19.
\bibitem{Meta} V. Metag, Nucl. Phys. {\bf A553} (1993) 283c.
\bibitem{KaRu} M. Kataja and P.V. Ruuskanen, Phys. Lett. {\bf B243}
(1990) 181.
\bibitem{LDHS} D. Lohse, J.W. Durso, K. Holinde and J. Speth,
Phys. Lett.  {\bf B234} (1989) 235;\newline
Nucl. Phys. {\bf A516} (1990) 513.
\bibitem{BbSu} R. Blankenbecler and R. Sugar, Phys. Rev.
{\bf 142} (1966) 1051.
\bibitem{PeHS} B. C. Pearce, K. Holinde and J. Speth, Nucl. Phys.
{\bf A541} (1992) 663.
\bibitem{ErWe} T. O. Ericson and W. Weise, 'Pions and Nuclei'
(Clarendon, Oxford 1988).
\bibitem{ChLo} G. F. Chew and F. E. Low, Phys. Rev.
{\bf 101} (1956) 1570.
\bibitem{BrWe} G. E. Brown and W. Weise, Phys. Rep.
{\bf C22} (1975) 221.
\bibitem{MSTV} A. B. Migdal, E. E. Saperstein, M. A. Troitsky and D. N.
Voskresensky, Phys. Rep. {\bf 192} (1990) 179.
\bibitem{Oste} F. Osterfeld, Rev. Mod. Phys. {\bf 64} (1992) 491.
\bibitem{aoui} Z. Aouissat, P. Schuck,
R. Rapp, J. Wambach and G. Chanfray, to be published.
\bibitem{Naga} S. Nagamiya, Nucl. Phys. {\bf A541} (1992) 5c.
\bibitem{KoBe} V. Koch and G. Bertsch, Nucl. Phys.
{\bf A552} (1993) 591.
\end{thebibliography}
\end{document}